\documentclass{iau} 

\usepackage{graphicx}
\usepackage{xcolor}
\usepackage[hyphens]{url}
\usepackage{hyperref}
\hypersetup{
     colorlinks  = true,
     linkcolor   = blue, 
     anchorcolor = BrickRed,
     citecolor   = blue,
     filecolor   = blue,
     menucolor   = blue,
     runcolor    = cyan,
     urlcolor    = blue
}
\usepackage[authoryear]{natbib}
\usepackage{epstopdf}	

\pubyear{2019}
\volume{355} 
\jname{The Realm of the Low-Surface-Brightness Universe}
\editors{D. Valls-Gabaud, I. Trujillo \& S. Okamoto, eds.}

\def\etal{\textit{et al.}}
\def\HI{H\,{\sc i}\, }
\def\Ms{$\textrm{M}_{\odot}$}
\def\kms{$\textrm{km~s$^{-1}$}$}

\title[ Unveiling the origin of gLSBs] 
{ Unveiling the origin of giant low surface brightness discs: \\ results of the long-slit spectral observations}

\author[Anna S. Saburova \etal]   
{Anna S. Saburova$^{1,2}$,
Igor V. Chilingarian$^{3,1}$,
Anastasia V. Kasparova$^{1}$,
Ivan Yu. Katkov$^{1, 4}$, 
Ol'ga K. Sil'chenko$^1$,
Roman I. Uklein$^5$}

\affiliation{$^1$ Sternberg Astronomical Institute, Moscow M.V. Lomonosov State University,\\ Universitetskij pr., 13,  Moscow, 119234, Russia\\ email: {\tt saburovaann@gmail.com} \\[\affilskip]
$^2$ Institute of Astronomy, Russian Academy of Sciences,\\ Pyatnitskaya st., 48, Moscow, 119017, Russia\\[\affilskip]
$^3$ Smithsonian Astrophysical Observatory,\\ 60 Garden Street MS09, Cambridge, MA 02138, USA\\[\affilskip]
$^4$ New York University Abu Dhabi,\\ Saadiyat Island, PO Box 129188, Abu Dhabi, UAE\\[\affilskip]
$^5$ Special Astrophysical Observatory, Russian Academy of Sciences,\\ Nizhniy Arkhyz, Karachai-Cherkessian Republic, 357147, Russia}

\begin{document}

\maketitle

\begin{abstract}
Giant low surface brightness galaxies (gLSB) with radii of discs as large as 130 kpc challenge galaxy formation scenarios and it is still not well understood how they form and evolve through the cosmic time. Here we present analysis of deep long-slit spectroscopic observations of six gLSBs that we  obtained with the Russian 6-m telescope: UGC 1922, Malin 2, UGC 6614, UGC1382, NGC 7589 and UGC 1378.   We derived spatially resolved properties of stellar and ionized gas kinematics and characteristics of stellar populations and interstellar medium. The stars in the central regions are old and metal rich for most of the galaxies. We revealed the presence of a kinematically decoupled central component in the inner regions of UGC1922, UGC1382 and UGC6614, where we detected counter-rotating kinematical components. We combine the results of our observations with the results available in literature. There seems to be a need for diversity of gLSBs formation scenarios: (i) some of them could have formed by in-plane mergers of massive galaxies; (ii) for some others the major merger scenario is excluded by our data. We revealed that most of gLSBs are situated in low-density environment which possibly helped to preserve the giant discs. At the same time at some point of the formation history of these systems there should exist a reservoir of gas from which the massive discs were formed.  Future observations and detailed comparison with numerical simulations of galaxy formation in the cosmological contest will help to clarify which gLSB formation channel is more important.
\keywords{galaxies: evolution, galaxies: formation, galaxies: structure, galaxies: individual (UGC1922, Malin2, Malin1, UGC 6614, UGC1382, NGC 7589, UGC 1378)}
\end{abstract}

\firstsection 

\section{Introduction}
Giant low surface brightness galaxies (gLSBs) represent a challenge for the current galaxy formation scenarios despite of being known for more than 30 years, since the prototype of this class of objects Malin~1 was discovered by \citet{Bothun1987}.  It is difficult to form such discs in the hierarchical clustering paradigm where dark haloes of disc galaxies do not undergo major mergers. Major merger events would likely destroy large
discs but merger trees of such massive galaxies with no major mergers are exceptionally rare. 

The sample of confirmed gLSBs stays limited despite of being extended as new
objects are discovered using improved observational facilities \citep[see, e.g.][]{Hagenetal2016}. It can indicate that gLSBs are rare and therefore a detailed insight on every object of this type helps to understand their nature.  That was the aim of our continuing series of works \citep[][]{Kasparova2014, Saburova2018, Saburova2019}.  We present the main properties of the sample of known gLSBs in Table \ref{properties} with the references to the origin of the data. The LSB disc radius gives 4 radial scale lengths of LSB discs for all galaxies except Malin~1 for which we used the distance to the last measured point 
above the noise level and with an approximate exponential radial
distribution of surface brightness given in \citet{Boissier2016}. All gLSBs demonstrate prominent spiral structure which continues at the low surface densities \citep[see, e.g.][]{Kasparova2014, Galaz2015, Hagenetal2016}. Many of them have complex structure when normal-size early type galaxy is embedded in LSB disc \citep[see, e.g.][]{Lelli2010, Saburova2019}.

We performed long-slit spectral observations of six gLSBs: UGC 1922, Malin 2, UGC 6614, UGC1382, NGC 7589 and UGC 1378 with the Russian 6-m telescope and the deep photometry of UGC~1922, UGC~1378 and Malin~2. For the details on the observations and data analysis see \citep[][] {Kasparova2014, Saburova2018, Saburova2019} and Saburova et al. in prep. We combined the results of our observations with the data available in literature including that for Malin~1 that was not a part of our observational sample to get clues on the formation and evolution of these unusual systems.

\section{On the results of the long-slit spectral analysis}
\begin{figure}[b]
\begin{center}
 \includegraphics[height=0.37\textwidth]{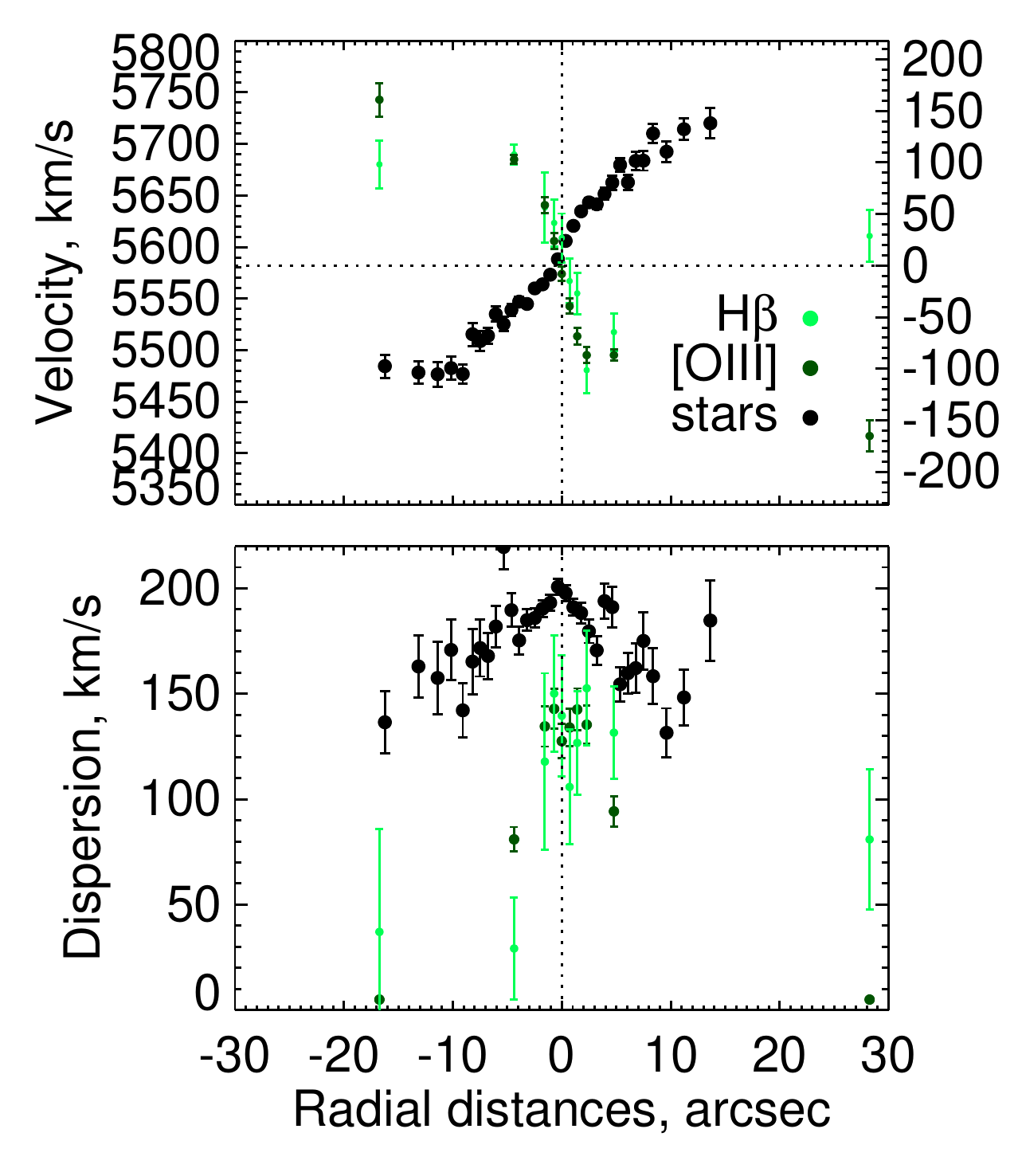} 
 \includegraphics[height=0.375\textwidth]{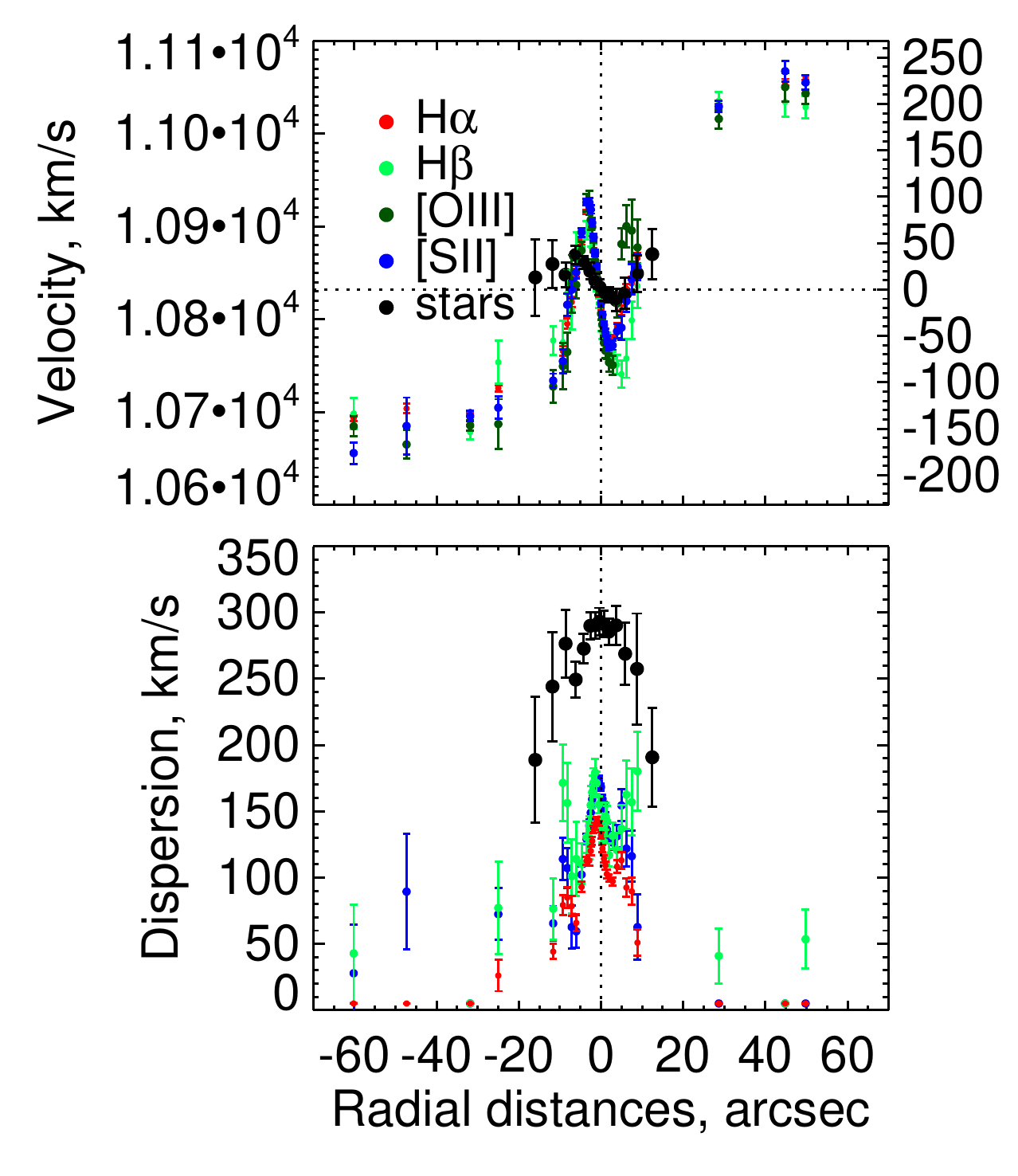} 
\includegraphics[height=0.37\textwidth]{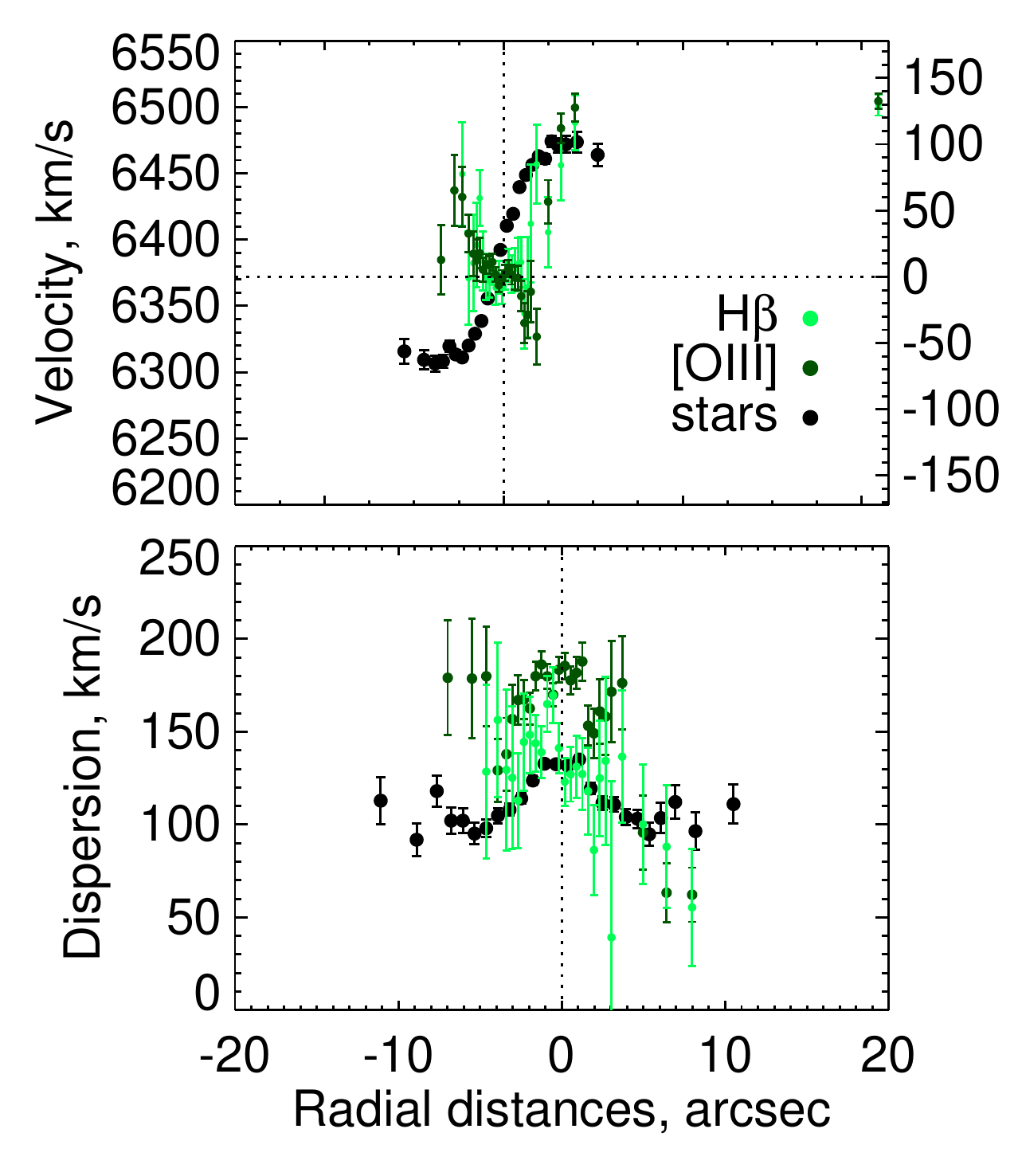} 
\caption{The radial profiles of velocity (top row) and velocity dispersion (bottom row) of ionized gas (colored symbols) and stars (black circles) for UGC~1382 ($PA=238^o$), UGC~1922 ($PA=85^o$) and UGC~6614 ($PA=45^o$) from left to right correspondingly.}
  \label{profiles}
\end{center}
\end{figure}

From our spectral observations we obtained spatially resolved properties of stellar and ionized gas kinematics and characteristics of stellar populations. We found that 3 of 6 considered gLSBs have counter-rotation of the gaseous disc with respect to the stellar disc or the outer gaseous disc. Similar high occurrence of decoupled gas kinematics was found in isolated S0 galaxies \citep{Katkov2014}. The stellar velocity profile obtained by \citet{Reshetnikov 2010} for the 7-th gLSB Malin~1 does not show counter-rotation with respect to outer \HI disc. The decoupled gaseous kinematics is demonstrated on the line-of-sight velocity profiles in Fig. \ref{profiles} (upper row). Ionized gas kinematics is shown by colored symbols and stellar kinematics is demonstrated by black points. We should admit however that we can not exclude that the peculiarity of the gaseous kinematics of UGC~6644 could be related to the outflow of gas related to the AGN. We need more spectral observations to make firm conclusion. The counter-rotation can indicate the presence of merger or accretion of gas in the history of these systems. We also found that the stellar population of the central parts of the gLSBs is very old ($T>10$ Gyr) and metal-rich ($[Z/H]=-0.2..0.2$ dex) in most cases besides galaxies with prominent bars UGC~1378 and NGC~7589.
\section{On the mass-modelling of the rotation curves}
\begin{figure}[b]
\begin{center}
\includegraphics[height=0.27\textwidth]{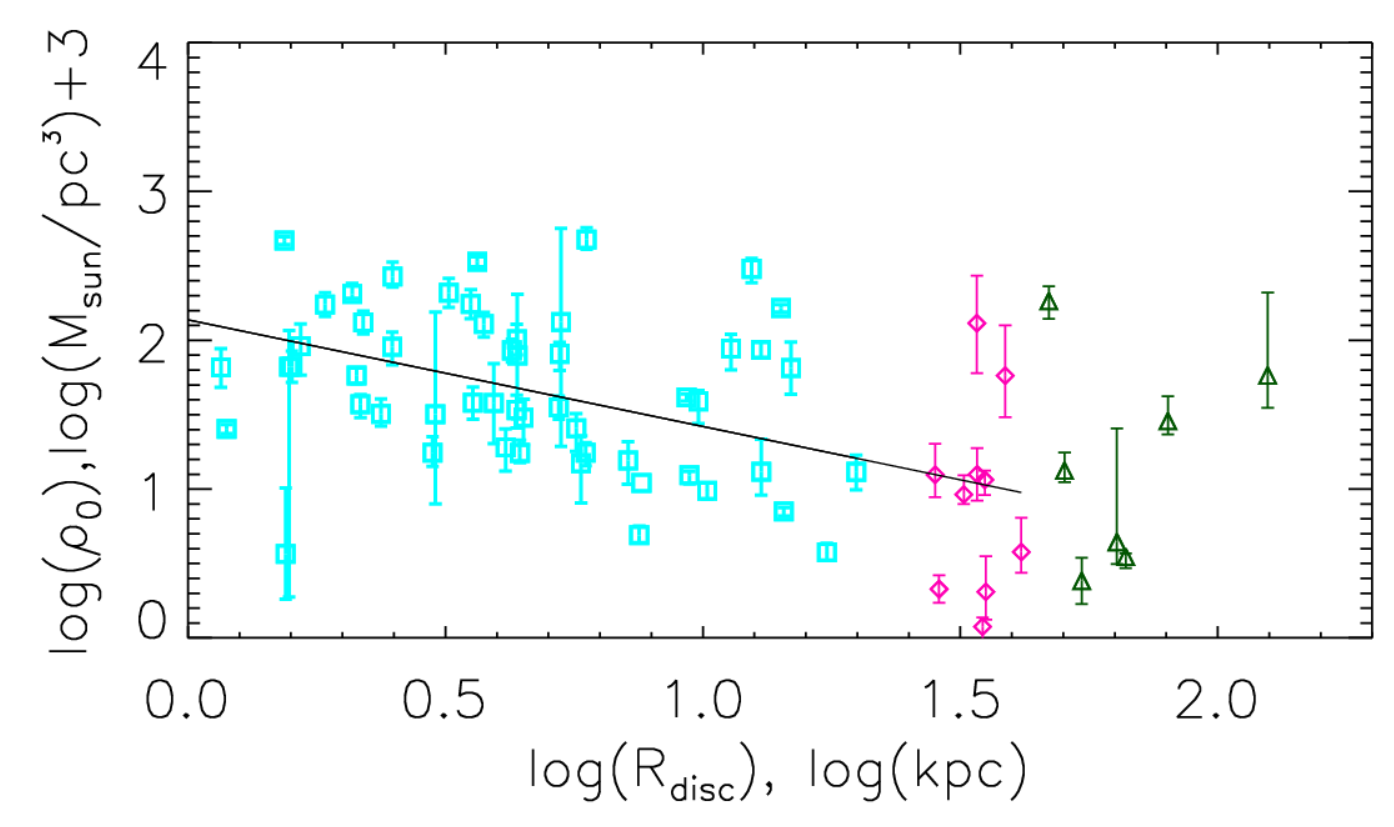} 
 \includegraphics[height=0.27\textwidth]{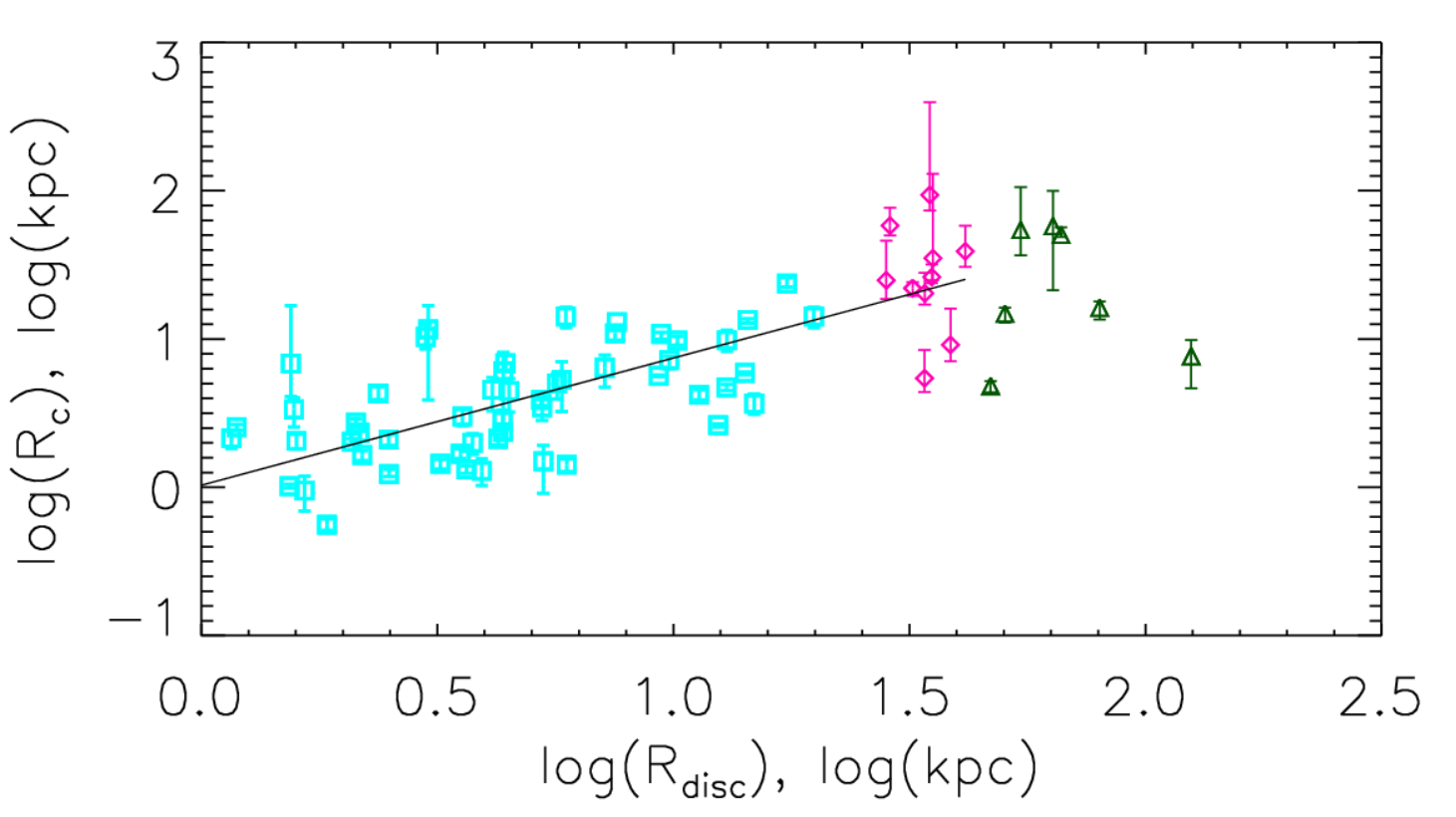} 
 \caption{The disc radius compared to the parameters of dark matter halo -- central density (left-hand panel) and radial scale (right-hand panel). The gLSBs are shown by triangles. Diamonds correspond to the giant HSBs, squares give the plot for moderate size galaxies, the line is the fit for HSB galaxies \citep{saburova}. }
  \label{fig_par}
\end{center}
\end{figure}

We combined the optical rotation curves obtained in the project with the \HI data from \citep{Lelli2010, Pickering1997, Mishra2017, Hagenetal2016}  and performed the mass-modelling of the combined rotation curves. We used the components of stellar and gaseous discs and bulges and \citep{ Burkert} dark halo. We fixed the densities of the stellar components using the multi-band photometrical data. The examples of the rotation curve decomposition could be found in \citep{Saburova2018, Saburova2019}, the details on the modeling are given in \citet{Saburova2016}. The mass-modelling allowed us to obtain the parameters of dark halo of gLSBs - the radial scale and the central density. We compared them to the radii of the discs in Fig. \ref{fig_par} (triangles), where we also demonstrate the parameters of HSB galaxies (squares and diamonds) \citep{saburova}.  The line corresponds to the least squares fit for HSB galaxies from \citet{saburova}. From Fig. \ref{fig_par} it follows that gLSBs behave differently on these plots. Some of them are lying on the continuation of the line plotted for HSB galaxies, other tend to deviate from it, like Malin~1 which dark halo radial scale is lower than expected for its high disc radius. It can indicate the different nature of gLSBs. 

\section{On the formation scenario for gLSBs}
The most important goal of our project was to understand how such systems as gLSBs could be formed. Here we summarize most of discussed formation scenarios. They can be divided into catastrophic and non-catastrophic. Among the \underline{catastrophic scenarios} are the following. (i) Bygone head on collision with massive intruder which forms a ring-like structure that evolves into the giant disc proposed by \citet{Mapelli2008}. This scenario seems to be unlikely because the deep images and color maps of the galaxies do not show the traces indicative for this model \citep[see e.g.][]{Boissier2016}. (ii) The formation of gLSB by tidally disrupted dwarf galaxies \citep{2006ApJ650L33P}. The major disadvantage of the scenario is that it predicts the decrease of the rotation velocity at the disc periphery which is not observed according to the \HI data \citep[see, e.g.][]{Mishra2017}, and also the satellites should have almost the same angular momentum to form disc not the spherical system, which makes this scenario less likely.  (iii) The group of scenarios in which an extended disc is formed from an ample supply of gas cooled down at the late stage of a merger \citep{Saburova2018, Zhu2018MNRAS}. These models show a good agreement with the observed properties of some of gLSBs, thus they seem to be more promising.  \underline{Non-catastrophic  scenarios} are the following.  
 (i) The formation of giant disc due  to the peculiarly high radial scale of dark halo. We can not exclude fully this scenario since most of the gLSB indeed have high radial scale of the halo \citep[see, e.g.][and Fig. \ref{fig_par}]{Kasparova2014, Saburova2018}. (ii) A build up
of a gLSB disc by accretion from cosmic filaments \citep{Saburova2018, Saburova2019}. This scenario has two stages. The first stage, in common with MW-type galaxies, may have included several episodes of merging and a second stage, following the accretion of most satellites, quiescently formed gLSB stellar and gaseous discs by accretion of metal-poor gas from a gas-rich filament.

\begin{figure}[b]
\begin{center}
 \includegraphics[height=0.4\textwidth]{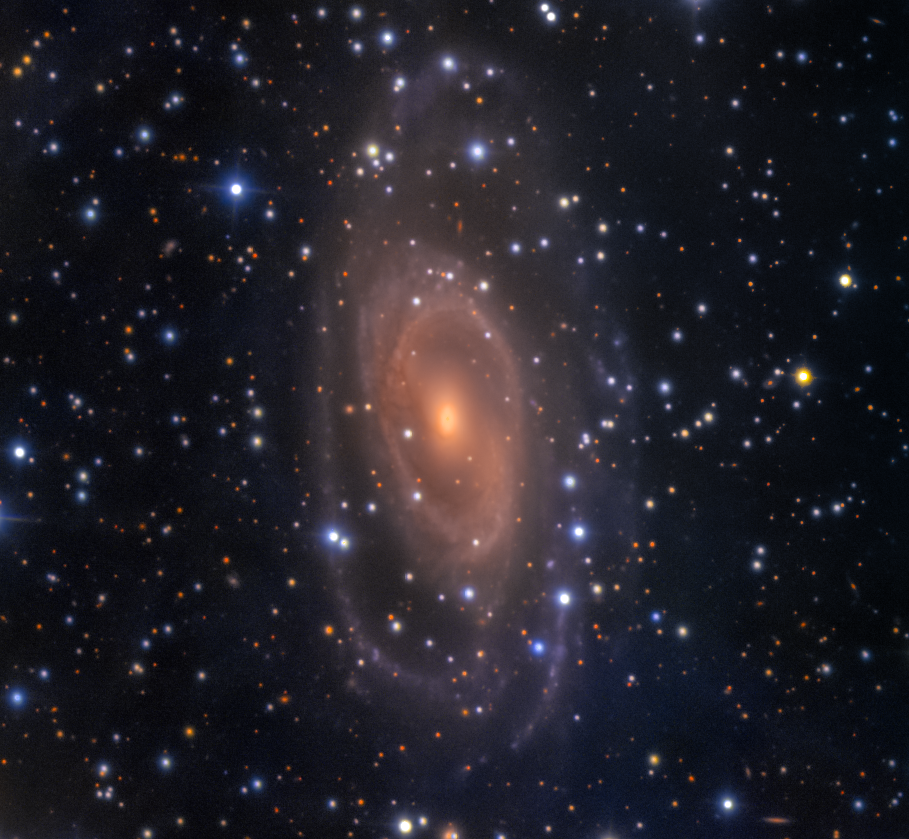} 
 \includegraphics[height=0.4\textwidth]{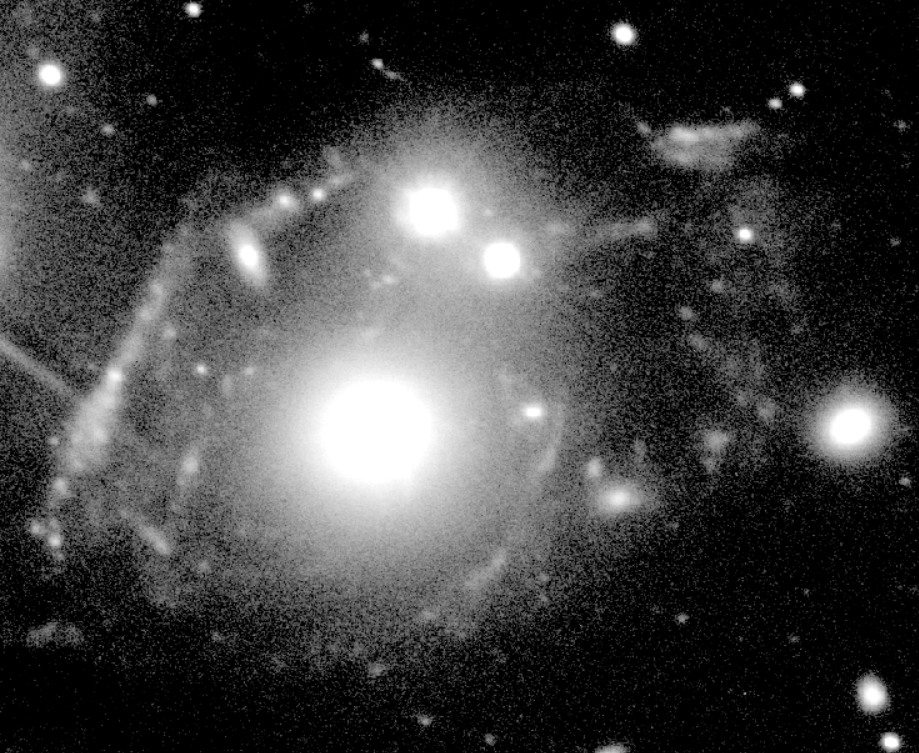} 
 \caption{Composite {\it r, g, z}-band image of UGC~1378 (left-hand panel) \citep{Saburova2019}, {\it g}-band image of UGC~1922 \citep{Saburova2018}.}
  \label{fig_images}
\end{center}
\end{figure}

According to our data there seems to be a need for diversity of gLSBs formation scenarios. The reasoning for it is visually demonstrated in  Fig. \ref{fig_images} where we show the deep images of two galaxies which both can be classified as gLSBs but which are completely different. On the left-hand side is \underline{UGC~1378} which has a bar and complex structure consisting of normal HSB galaxy with bulge and disc. According to our estimates of the stellar velocity dispersion, its disc is not dynamically overheated which is expected in major merger scenario \citep{Saburova2019}. We do not see any peculiarities in its morphology and kinematics. It is located in a low-density
environment \citep{Saburova2019}.

On the right-hand is \underline{UGC~1922} which seems to be a bulge embedded in the LSB disc, and has peculiar morphology. Its stellar velocity dispersion is high and corresponds to the strongly overheated disc. We also found counter-rotation of the outer gaseous disc with respect to the inner one (see Fig. \ref{profiles}). The radius of the disc of UGC~1922 is also higher than that of UGC~1378 (see Table \ref{properties}). UGC~1922 belongs to a group
that includes 7 spectroscopically confirmed members, one of which
is a giant elliptical galaxy that dominates the group \citep{Saburova2018}.

 Most likely these two objects were formed differently. First one was formed by the gas accretion on the HSB galaxy. And the second one is better described by major merger scenario.

\begin{table}
  \begin{center}
  \caption{The main properties of gLSBs.}
  \label{properties}
 {\scriptsize
  \begin{tabular}{|l|c|c|c|c|c|c|c|}\hline 
{\bf Galaxy}&Ra, Dec&{\bf LSB disc radius}&{\bf Type}&{\bf \HI mass}&{\bf Distance}&{\bf Inclination}&{\bf Rotation velocity}\\
&&(kpc)&&($10^{10}$\Ms) &(Mpc)&($^o$)&(\kms)\\
\hline
Malin~1&12h36m59.350s, &130$^2$&SBab$^3$ &6.7$^4$&377$^4$&38$^{4}$&236$^5$\\
&+14d19m49.32s$^1$&& &&&&\\
\hline
Malin~2&10h39m52.483s, &82$^6$&Scd$^3$&3.6$^7$&201$^8$&38$^7$&320$^7$\\
&+20d50m49.36s$^1$&& &&&&\\
\hline
UGC~6614&11h39m14.872s, &54$^7$&Sa$^3$&2.5$^7$&85$^7$&35$^7$&228$^7$\\
&+17d08m37.21s$^1$&& &&&&\\
\hline
NGC~7589&23h18m15.668s, &56$^4$&SABa$^3$&1.5$^4$&130$^4$&58$^7$&205$^4$\\
&+00d15m40.19s$^1$&& &&&&\\
\hline
UGC~1378&01h56m19.24s, &50$^9$&SBa$^3$&1.2$^{10}$&38.8$^{10}$&59$^{10}$&280$^{10}$\\
&+73d16m58.0s$^1$&& &&&&\\
\hline
UGC~1382&01h54m41.042s, &80$^{11}$&S0$^{11}$&1.7$^{11}$&80$^{11}$&46$^{11}$&280$^{11}$\\
&-00d08m36.03s$^1$&& &&&&\\
\hline
UGC~1922&02h27m45.930s, &84$^{12}$&S?$^1$&3.2$^{10}$&150$^{10}$&51$^{10}$&432$^{10}$\\
&+28d12m31.83s$^1$&& &&&&\\
\hline
  \end{tabular}
  }
 \end{center}
\vspace{1mm}
 \scriptsize{
 {\it References:}\\
 $^1$ NED \footnote{http://ned.ipac.caltech.edu}
  $^2$ \citet{Boissier2016},
  $^3$ Hyperleda database\footnote{http://leda.univ-lyon1.fr/}, \citet{Makarovetal2014},
$^4$ \citet{Lelli2010},
$^5$ \citet{Moore2006},
$^6$ \citet{Kasparova2014},
$^7$ \citet{Pickering1997}
$^8$ \citet{Das2010}
$^9$ \citet{Saburova2019}
$^{10}$ \citet{Mishra2017}
$^{11}$ \citet{Hagenetal2016}
$^{12}$ \citet{Saburova2018}}
\end{table}

We analyzed each galaxy of the sample in the similar way and decided which of the considered scenarios is more likely for it. We discuss it below. 

\underline {Malin~1}. We can rule out the scenario in which the giant size of the disc is due to the peculiarly high radial scale of the dark matter halo, since it contradicts our estimates, \citet{Lelli2010} also give moderate value of the radial scale of the dark halo for Malin~1. The presence of a complex structure with two discs makes the scenario of accretion of the gas from the filament on the preformed early-type galaxy more possible for this system. LSB disc could be the result of the accretion.  \citet{Zhu2018MNRAS} proposed a major merging scenario for this galaxy which also does not contradict its observed properties including the absence of the gradient of the stellar age in its disc. However, to make firm conclusion one needs to estimate the stellar velocity dispersion in the region of LSB disc of the galaxy.\\
\underline {Malin~2}. We can not exclude the possibility that extended disc of Malin~2 is due to the high radial scale of the dark halo \citep[in agreement with our finding in][]{Kasparova2014}. The major merger scenario is less likely since the stellar velocity dispersion is not very high at one disc scalelength and we observe high gradient of gas metallicity in the disc  \citep{Kasparova2014}. The formation of the giant disc from the gas accreted from the filament could not be fully excluded.\\
\underline{UGC~6614}. For this galaxy we need more information to choose the scenario (properties of stellar population and gas metallicty in the region of LSB-disc). We can not exclude both catastrophic and non-catastrophic scenarios. \\
\underline{NGC~7589} possesses massive companion NGC~7603 which has disturbed spiral arms. Both extended disc of NGC~7589 and the disturbed appearance of NGC~7603 could be the traces of interaction of these two galaxies. The gas accretion scenario however could not be fully excluded since NGC~7589 has HSB and LSB discs \citep{Lelli2010}. \\
\underline{UGC~1382}. The counter-rotation of gaseous disc could indicate the accretion of the gas on the pre-formed early-type galaxy. \HI velocity field is not disturbed \citep{Hagenetal2016} which gives evidences against recent major merger. We also can rule out the scenario in which the extended disc is formed due to unusually high radial scale of dark halo, since the parameters of the halo deviate from the expected for high radius of the LSB-disc. 
\begin{discussion}
\discuss{G. Galaz}{Two questions: (1) In the case of Malin 1, the presence of Malin 1B and a stream show an 
indication of a possible interaction in the past even in a low-density environment. (2) Do you have collision
models involving Sc and Sd galaxies?}
\discuss{A. Saburova}{(1) Yes, indeed, interactions are unavoidable even in low-density environment. Malin~2 and UGC~1382 also demonstrate signs of minor merger. However it could be that the number of interactions is lower in the sparse environment that could help to preserve the giant discs. (2) We performed N-body/hydrodynamical simulations of in-plane merger of giant Sa and Sd galaxies and got giant LSB galaxy as a result. It can indicate that some of giant LSB galaxies could be formed by major merger. }

\discuss{A. Watkins}{The presence of bars and rings makes some of these galaxies look normal but for their, 
sheer size. Is it just a matter of different outskirts?}
\discuss{A. Saburova}{Yes, many giant LSB galaxies demonstrate a complex structure consisting of normal high surface brightness galaxy surrounded by giant disc of low surface brightness. }

\end{discussion}

\section{Acknowledgements} 
AS is grateful to IAU Grant for financial support to attend an IAU Symposium 355.
AS acknowledges  The Russian Science Foundation (RSCF) grant  19-72-20089  that supported research on dynamical modelling of gLSBs. IC and AS acknowledge The Russian Science Foundation (RSCF) grant 19-12-00281 for the reduction and analysis of spectral and photometric data.

\end{document}